
%
%
%
\normalbaselineskip = 12 pt
\magnification = 1200
\hsize = 15 truecm \vsize = 22 truecm \hoffset = 1.0 truecm
\rightline{LANDAU-92-TMP-2}
\rightline{December 1992}
\rightline{Submitted to Mod. Phys. Lett. A}
\medskip
\centerline{{\bf SUPERCONFORMAL 2D MINIMAL MODELS AND AN UNUSUAL}}
\medskip
\centerline{{\bf COSET CONSTRUCTION}}
\medskip
\centerline{M. YU. LASHKEVICH\footnote{$^*$}{ This work was supported,
in part, by Landau Scolarship
Grant awarded by Forschungszentrum J\"ulich GmbH, and by Soros Foundation
Grant awarded by the American Physical Society}}
\medskip
\centerline{ Landau Institute for Theoretical Physics, Academy of Sciences, }
\centerline{ Kosygina 2, GSP-1, 117940 Moscow V-334, Russian Federation
\footnote{$^{**}$}{E-mail: ngm@cpuv1.net.kiae.su with note
`Lashkevich/Landau Inst.' in Subject.}
}
\medskip
\par
\noindent
We consider a coset construction of minimal models. We define it rigorously
and prove that it gives superconformal minimal models. This construction
allows to build all primary fields of superconformal models and
to calculate their three-point correlation functions.
\bigskip
\par\noindent
Superconformal two-dimensional unitary minimal models$^{1-4}$ are theories with
chiral algebra which is generated by energy-momentum tensor $T^S(z)$ of spin
$2$ and its superpartner $J(z)$ of spin ${3\over 2}$. Their operator
product expansions are given by
$$\eqalignno{
&T^S(z')T^S(z)={{1\over 2}c^S\over(z'-z)^4}+{2T^S(z)\over(z'-z)^2}
+{\partial T^S(z)
\over z'-z}+O(1),&(1{\rm a})\cr
&T^S(z')J(z)={{3\over 2}J(z)\over(z'-z)^2}+{\partial J(z)\over z'-z}+O(1),
&(1{\rm b})\cr
&J(z')J(z)={{3\over 2}c^S\over(z'-z)^3}+{2T^S(z)\over z'-z}
+\partial T^S(z)+O(z'-z),&(1{\rm c})}
$$
with $\partial\equiv\partial/\partial z$, and $c^S$
be a central charge of the Virasoro
algebra (1a)\footnote{$^{\rm a}$}{We use
the definition of central charge common for conformal models.
It is ${3\over 2}$ of superconformal one.}
$$
c_k^S={3\over 2}-{12\over(k+2)(k+4)};\ k=1,2,3,\cdots.\eqno(2)
$$
We will designate these models as SM$_k$. It is known$^5$ that superconformal
minimal models can be obtained as a result of coset construction of
SU$(2)$ Wess-Zumino models:
$$
{\rm SM}_k\sim{{\rm SU}(2)_k\times{\rm SU}(2)_2\over{\rm SU}(2)_{k+2}}.
\eqno(3)
$$
Index at SU$(2)$ designates central charge of Kac-Moody algebra.$^6$
On the other hand, conformal minimal model M$_k$ with central charge
$$
c_k=1-{6\over(k+2)(k+3)}\eqno(4)
$$
can be obtained by coset construction too
$$
{\rm M}_k\sim{{\rm SU}(2)_k\times{\rm SU}(2)_1\over{\rm SU}(2)_{k+1}}.
\eqno(5)
$$
Let us make a naive transformation
$$\eqalign{
{\rm SM}_k&\sim{{\rm SU}(2)_k\times{\rm SU}(2)_1\over{\rm SU}(2)_{k+1}}\times
{{\rm SU}(2)_{k+1}\times{\rm SU}(2)_1\over{\rm SU}(2)_{k+2}}\times
{{\rm SU}(2)_2\over{\rm SU}(2)_1\times{\rm SU}(2)_1}\cr
&\sim{{\rm M}_k\times{\rm M}_{k+1}\over{\rm M}_1}.}\eqno(6)
$$
The last expression is quite unusual. Can we assign an exact meaning to it?
Before we make it we mention a generalization to coset constructions
$$
{\rm N}_{kl}\sim{{\rm SU}(2)_k\times{\rm SU}(2)_l\over{\rm SU}(2)_{k+l}}.
In the same way as in (6) we can write
$$
$$
{\rm N}_{kl}\sim{{\rm N}_{k,l-m}\times{\rm N}_{k+l-m,m}
\over{\rm N}_{l-m,m}},
$$
in particular
$$
{\rm N}_{kl}\sim{{\rm N}_{k,l-1}\times{\rm M}_{k+l-1}
\over{\rm M}_{l-1}}\sim{{\rm M}_k\times{\rm N}_{k+1,l-1}
\over{\rm M}_{l-1}}.
$$
We defer consideration of this general case to a more detailed paper.
\par
Now we will look at the representation of minimal model M$_k$ by one bosonic
field $\varphi(z)^{7-10}$ with correlation function
$$
\langle\varphi(z')\varphi(z)\rangle=-\ln(z'-z),
$$
and energy-momentum tensor
$$
T_k(z)=-{1\over 2}:(\partial\varphi)^2:+{i\over\sqrt{2(k+2)(k+3)}}
\partial^2\varphi,
$$
colons mean normal ordering. Vertex operators, $V^{(k)mn}_{\ \ \ (p,q)}(z)$,
with conformal dimensions
$$
\Delta_{(p,q)}^{(k)}={[(k+3)p-(k+2)q]^2-1\over 4(k+2)(k+3)}
$$
are given by
$$\eqalign{
&V_{\ \ \ (p,q)}^{(k)mn}(z)=V_{(p,q)}^{(k)}(z)
\prod_{i=1}^m\oint_{C_i}du_iI_+^{(k)}(u_i)
\prod_{i=1}^n\oint_{C_i}dv_iI_-^{(k)}(v_i),\cr
&\eqalign{
V_{(p,q)}^{(k)}(z)&=:\exp\left(-i\left(\textstyle
{{p-1\over 2}\sqrt{2{k+3\over k+2}}
-{q-1\over 2}\sqrt{2{k+2\over k+3}}}\right)\varphi(z)\right):,\cr
&p=1,\cdots,k+1;\ \ q=1,\cdots,k+2;}}\eqno(7)
$$
where screenings,
$$\eqalign{
&I_+^{(k)}(z)=:\exp\left(i\textstyle{\sqrt{2{k+3\over k+2}}}\varphi\right):,\cr
&I_-^{(k)}(z)=:\exp\left(-i\textstyle{\sqrt{2{k+2\over k+3}}}\varphi\right):,}
\eqno(8)
$$
are fields with conformal dimension $1$; contours of integration
are presented in
\noindent
Fig. 1. Recall that vertex operators form an irreducible
representation for quantum group$^{11}$
SL$_{q(k)}(2)\odot$SL$_{\overline{q(k+1)}}(2)$,
$q(k)=\exp\left({2\pi i\over k+2}\right)$; bar means complex conjugation;
sign $\odot$ means a special
kind of product.$^{11,12\ }$\footnote{$^{\rm b}$}{The origin of this product
is in hidden SL$_{q(1)}(2)$ group which corresponds to the monodromy
trivial situation. Full quantum group of minimal model M$_k$ is$^{15}$
SL$_{q(k)}(2)\times$SL$_{q(1)}(2)\times$SL$_{\overline{q(k+1)}}(2)$
[see also Eq. (5)].}
This representation have `full momenta' $J_1={1\over 2}(p-1)$,
$J_2={1\over 2}(q-1)$ for two SL$(2)$ groups. `Projections of momenta'
are $M_1=J_1-m$, $M_2=J_2-n$.
\par
 From two representations with equal dimensions,
the first for group SL$_q(2)$ and the second for SL$_{\overline{q}}(2)$,
a bilinear braiding ($R$-matrix) invariant can be constructed.
Therefore, invariant
primary field can be constructed as
$$
\phi_{(p,q)}^{(k)}(z,\overline{z})=N_{(p,q)}\sum_{m=0}^{p-1}\sum_{n=0}^{q_1}
X_p^{(k)}(m)X_q^{(k+1)}(n)V_{\ \ \ (p,q)}^{(k)mn}(z)
V_{\ \ \ (p,q)}^{(k)mn}(\overline{z}),
$$
where $X_p^{(k)}(m)$ are the coefficients of invariants for
SL$_{q(k)}(2)$ quantum group, and $N_{(p,q)}$ is a normalization factor
which will be omitted later. Notice that this braiding invariance provides
monodromy invariance$^{7,8}$ or, equivalently, duality$^{13}$ of
correlation functions.
\par
Let us construct a vertex operator
$$
W_{(p,s,q)}^{mn}(z)=\sum_{r=0}^{s-1}X_s^{(k+1)}(r)
V_{\ \ \ (p,s)}^{(k)mr}(z)V_{\ \ \ \ \ \ (s,q)}^{(k+1)rn}(z)\eqno(9)
$$
in the theory M$_k\times$M$_{k+1}$. This vertex realizes a represetation of
the quantum group SL$_{q(k)}(2)\times$SL$_{\overline{q(k+2)}}(2)$.
We have constructed the numerator of
the coset (6). Now we will build the denominator.
\par
The superconformal minimal model SM$_k$ possesses a representation
by three bosons$^{14}$ $\chi(z)$, $\rho(z)$ and $\Phi(z)$ with non-zero
correlation functions
$$
\langle\chi(z')\chi(z)\rangle=\langle\rho(z')\rho(z)\rangle=
\langle\Phi(z')\Phi(z)\rangle=-\ln(z'-z),
$$
and energy-momentum tensor
$$\eqalign{
T_k^S(z)=-{1\over 2}(\partial\chi)^2-{i\over 2}\partial^2\chi
-{1\over 2}(\partial\rho)^2+&{1\over 2\sqrt{2}}\partial^2\rho\cr
-{1\over 2}(\partial\Phi)^2+&{i\over\sqrt{(k+2)(k+4)}}\partial^2\Phi.}
$$
The supercurrent is given by (up to a normalization factor)
$$\eqalign{
J(z)\sim\oint_{C_z}{du\over 2\pi i}:\exp&\left(i(k+1)\chi
-(k+2)\sqrt{2}\rho-i\sqrt{(k+2)(k+4)}\Phi\right):(u)
\cr
&\cdot:\exp\left(-ik\chi+{2k+3\over\sqrt{2}}\rho
+i\sqrt{(k+2)(k+4)}\Phi\right):(z).}
$$
Vertex operators are given by
$$\eqalign{
S_{\ \ \ \ pp'q}^{(k)mm'n}(z)&=S_{pp'q}^{(k)}(z)
\cr
&\cdot
\prod_{i=1}^m\oint_{C_i}du_iI_1(u_i)
\prod_{i=1}^{m'}\oint_{C'_i}dv_iI'_1(v_i)
\prod_{i=1}^{n}\oint_{S_i}dw_iI_2(w_i),\cr
S_{pp'q}^{(k)}&=f_{p+p'-q-1+4N}(z):\exp\Bigl\{
\textstyle{2(q-p-4N)-(p'-1)\over 2\sqrt{2}}
\rho
\cr
&\ \ \ \ \ \ \ \
-i\textstyle{(k+4)(p-1)-(k+2)(q-1)\over 2\sqrt{(k+2)(k+4)}}\Phi\Bigr\}:(z),
\cr
&p=1,\cdots,k+1; \ \ p'=1,2,3;\ \ q=1,\cdots,k+3;\cr
&p+p'-q-1\in 2{\bf Z},\ \ (1,3,1)\neq(p,p',q)\neq(k+1,1,k+4),\cr
&0\leq q+p'-p-1-4N<4,\ \ N\in{\bf Z},\cr
&f_{2n}(z)=\cases{:e^{in\chi}:(z) & if $n\geq 0$,\cr
\oint_{C_z}{du\over 2\pi i}:e^{i\chi}:(u):e^{-i(1-n)\chi}:(z) & if
$n\leq 0$,}
}\eqno(10)
$$
contours $C_i$, $C'_i$ and $S_i$ are as in Fig. 1, contour $C_z$ is a
small circle around $z$. Screening fields are given by\footnote{$^{\rm c}$}
{In Ref. 14 they are designated as $I_1\tilde{Z}$, $I'_1$ and $I_2Z$
correspondingly.}
$$\eqalign{
&I_1(z)=\oint_{C_z}{du\over 2\pi i}:e^{3i\chi-2\sqrt{2}\rho}:(u)
:\exp\left(-2i\chi+\sqrt{2}\rho
+i\textstyle{\sqrt{k+4\over k+2}}\Phi\right):(z),
\cr
&I'_1(z)=\oint_{C_z}{du\over 2\pi i}:e^{i\chi}:(u)
:\exp\left(-2i\chi+\textstyle{1\over\sqrt{2}}\rho\right):(z),
\cr
&I_2(z)=\oint_{C_z}{du\over 2\pi i}:e^{i\chi}:(u)
:\exp\left(-2i\chi+\sqrt{2}\rho
-i\textstyle{\sqrt{k+2\over k+4}}\Phi\right):(z).
}\eqno(11)
$$
Note that the supercurrent $J(z)$ does not commute with the screening operator
\noindent
$\oint duI'_1(u)$. It can be proved, however, that $J(z)$ exhibits trivial
monodromy properties.$^{15}$ Indeed, using Dotsenko procedure$^{16}$
we obtain immidiately the fusion rule
$$
J(z')\psi_{pp'q}^{(k)}(z)\sim(z'-z)^{-3/2\pm 1/2}
\left[\psi_{p,4-p',q}^{(k)}\right],
\eqno(12)
$$
where $\psi_{pp'q}^{(k)}$ is the field described by vertex
$S_{\ \ \ pp'q}^{(k)mm'n}(z)$, brackets mean the field with its entire
Fock module. This means that, if $z'$ goes around $z$ and returns to
the initial point, the product only changes by a constant factor
(here unity), and any correlation function with chiral (holomorphic)
current $J(z)$ is well defined.
\par
Conformal dimensions of vertices (10) are given by$^{2-4,14}$
$$
\Delta_{pp'q}^{(k)}={[(k+4)p-(k+2)q]^2-4\over 8(k+2)(k+4)},\eqno(13{\rm a})
$$
if $p-q\in 2{\bf Z}$, $p'-1=p-q\ ({\rm mod}\ 4)$ (Neveu-Schwarz primaries),
$$
\Delta_{pp'q}^{(k)}={[(k+4)p-(k+2)q]^2-4\over 8(k+2)(k+4)}+{1\over 2},
\eqno(13{\rm b})
$$
if $p-q\in 2{\bf Z}$, $p'-3=p-q\ ({\rm mod}\ 4)$ (Neveu-Schwarz superpartners),
$$
\Delta_{pp'q}^{(k)}={[(k+4)p-(k+2)q]^2-4\over 8(k+2)(k+4)}+{1\over 16},
\eqno(13{\rm c})
$$
if $p-q\in 2{\bf Z}+1$ (Ramond primaries).
\par
Quantum group of the superconformal minimal model is
SL$_{q(k)}(2)\times$SL$_{q(2)}(2)$ $\times$SL$_{\overline{q(k+2)}}(2)$,
and invariant primary fields are
$$
\psi_{pp'q}^{(k)}(z,\overline{z})=\sum_{mm'n}X_p^{(k)}(m)
X_{p'}^{(2)}(m')X_q^{(k+2)}(n)S_{\ \ pp'q}^{(k)mm'n}(z)
S_{\ \ pp'q}^{(k)mm'n}(\overline{z}).
$$
In Neveu-Schwarz case monodromy invariant fields can be obtained
without summation over $m'$ by the same reason as $J(z)$
is a chiral current [see Eq. (12)], and the following expression is
already monodromy invariant:
$$
\psi_{pp'p''q}^{(k)\prime}(z,\overline{z})=\sum_{mn}X_p^{(k)}(m)
X_q^{(k+2)}(n)S_{\ \ \ \ pp'q}^{(k)mn}(z)
S_{\ \ \ \ pp''q}^{(k)mn}(\overline{z}).
$$
Here absence of $m'$ in vertex means that necessary summation is
already done in each vertex separately.
\par
Consider now a vertex in M$_2\times$SM$_k$ theory
$$
U_{pp'q,t}^{mn}(z)=\sum_{m'=1}^{p'-1}X_{p'}^{(2)}(m')
V_{\ \ \ (t,p')}^{(1)rm'}(z)S_{\ \ \ \ pp'q}^{(k)mm'n}(z),\ \ \ t=1,2.\eqno(14)
$$
Index $r$ corresponds to SL$_{q(1)}(2)$ quantum group which describes
monodromy trivial situation, and we omit it in the l.h.s.
Let us prove that the theory
generated by vertices $W_{(p,s,q)}^{mn}(z)$ from Eq. (9) contains
the theory generated by vertices $U_{pp'q,t}^{mn}(z)$. We will apply the
method used in the proof of equivalence of coset theory
SU$(2)_k\times$SU$(2)_l/$SU$(2)_{k+l}$ and minimal-like bosonic
models.$^{15}$ Namely, we construct chiral currents $T_k(z)$ and
$T_{k+1}(z)$ of M$_k\times$M$_{k+1}$ theory from vertices (14) of
M$_1\times$SM$_k$ theory:
$$\eqalign{
&T_k(z)=\textstyle{{1\over 2}{k\over k+3}}T_1(z)
+\textstyle{{1\over 2}{k+4\over k+3}}T_k^S(z)
+\textstyle{\sqrt{(k+2)(k+4)}\over 2(k+3)}J(z)\phi_{(2,1)}^{(1)\prime}(z),
\cr
&T_{k+1}(z)=\textstyle{{1\over 2}{k+6\over k+3}}T_1(z)
+\textstyle{{1\over 2}{k+2\over k+3}}T_k^S(z)
-\textstyle{\sqrt{(k+2)(k+4)}\over 2(k+3)}J(z)\phi_{(2,1)}^{(1)\prime}(z),
}\eqno(15)
$$
$$
T_k(z)+T_{k+1}(z)=T_1(z)+T_k^S(z),\eqno(16)
$$
where $\phi_{(2,1)}^{(1)\prime}(z)$ is a chiral field in M$_1$. M$_1$ with
such chiral field is known to coinside with fermion theory.$^{1,13,17}$
The field $\phi_{(2,1)}^{(1)\prime}(z)$ is the fermion current
with spin ${1\over 2}$ and operator product expansion
$$
\phi_{(2,1)}^{(1)\prime}(z')\phi_{(2,1)}^{(1)\prime}(z)
=(z'-z)^{-1}+2(z'-z)T_1(z)+(z'-z)^2\partial T_1(z)
+O\left((z'-z)^3\right).\eqno(17)
$$
Correct operator product expansions for $T_k(z)$ and $T_{k+1}(z)$
from Eq. (26)
are directly obtained from Eqs. (1) and (17). These energy-momentum
tensors act locally on vertices $U_{pp'q,t}^{mn}(z)$, and therefore,
these vertices belong to M$_k\times$M$_{k+1}$. Quantum group
properties show us that these vertices belong to the subtheory generated by
vertices $W_{(p,s,q)}^{mn}(z)$.
\par
Note that one can uniquely determine conformal blocks of
$S_{\ \ \ \ pp'q}^{(k)mm'n}(z)$ if those of
$U_{pp'q,t}^{mn}(z)$ and $V_{\ \ \ (t,p')}^{(2)rm'}(z)$
are known.$^{18}$ Indeed, matrix $V_i^j=\langle V_{\ \ (t,p')}^{(2)rm'}
\cdots\rangle$, $i=(t,p',...)$, $j=(r,m,...)$, truncated in evident way
is indegenerate, because its
determinant is Wronskian of linearly independent solutions of those
differential equations which define conformal theory M$_1$.
\par
Let us write out $T_1(z)$, $T_k^S(z)$, and $J(z)$ through
M$_k\times$M$_{k+1}$ fields. Consider the field
$$
t(z)=W_{(1,3,1)}(z).\eqno(18)
$$
Its monodromy properties are trivial and its operator product expansion
can be calculated using bosonic representations for M$_k$ and M$_{k+1}$
in just the same way as tree-point correlation functions for
minimal models.$^{8,9}$ Instead of gluing together holomorphic and
antiholomorphic parts in calculation of correlation functions we must
glue together M$_k$ and M$_{k+1}$ parts of the field. The result is
$$\eqalign{
&t(z')t(z)={1\over(z'-z)^4}+{2\theta(z)\over(z'-z)^2}
+{\partial\theta(z)\over z'-z}+O(1),
\cr
&\eqalign{
\theta(z)=\textstyle{2(k-1)(k+2)\over\sqrt{3k(k+1)(k+5)(k+6)}}t(z)
&+\textstyle{(k+1)(k+2)\over k(k+5)}T_k(z)
\cr
&+\textstyle{(k+4)(k+5)\over(k+1)(k+6)}T_{k+1}(z).
}}\eqno(19)
$$
Finally, we have
$$\eqalign{
&T_1(z)=\textstyle{{1\over 4}{k+2\over k+5}}T_k(z)
+\textstyle{{1\over 4}{k+4\over k+1}}T_{k+1}(z)
+\textstyle{{1\over 4}\sqrt{3{k(k+6)\over(k+1)(k+5)}}}t(z),
\cr
&T_k^S(z)=\textstyle{{3\over 4}{k+6\over k+5}}T_k(z)
+\textstyle{{3\over 4}{k\over k+1}}T_{k+1}(z)
-\textstyle{{1\over 4}\sqrt{3{k(k+6)\over(k+1)(k+5)}}}t(z),}\eqno(20)
$$
$$\eqalign{
\phi_{(2,1)}^{(1)\prime}(z)J(z)=\textstyle{
\sqrt{3{k(k+6)\over(k+1)(k+2)(k+3)(k+5)}}}
t(z)&+\textstyle{{1\over 2}{(k+2)(k+6)\over(k+3)(k+5)}}T_k(z)
\cr
&-\textstyle{{1\over 2}{k(k+4)\over(k+1)(k+3)}}T_{k+1}(z).}\eqno(21)
$$
\par
Now we can make field identification. Consider operator product
expansion
$$\eqalign{
t(z')W_{(p,s,q)}(z)\sim(z'-z)^{-2}[W_{(p,s,q)}]
&+(z'-z)^{-2s+p+q}[W_{(p,s-2,q)}]
\cr
&+(z'-z)^{2s-p-q}[W_{(p,s+2,q)}].}
$$
Fields $W_{(p,s\pm 2,q)}$ do not give poles of the order $>2$ if
$$
\textstyle{1\over 2}(p+q)-1\leq s\leq\textstyle{1\over 2}(p+q)+1.
\eqno(22)
$$
This means that, if the condition (22) is satisfied, the fields
$W_{(p,s,q)}(z)$ can be primary. We will discuss all cases in sequence.
\par
$\underline{1.\ p+q\in 2{\bf Z},\ s={1\over 2}(p+q).}$ Considering
operator product expansions of
\par\noindent
$(T_k(z')+T_{k+1}(z'))W_{(p,s,q)}(z)$
and $T_k^S(z')W_{(p,s,q)}(z)$ and using Dotsenko-Fateev tech\-ni\-que$^{8,9}$
we calculate conformal dimensions and see that we have Neveu-Schwarz
primary fields
$$
S_{\ \ \ pp'q}^{(k)mn}(z)=W_{(p,s,q)}^{mn}(z),\ \
s=\textstyle{1\over 2}(p+q),\ \ p'-1=p-q\ ({\rm mod}\ 4).\eqno(23)
$$
\par
$\underline{2.\ p+q\in 2{\bf Z}+1,\ s={1\over 2}(p+q\pm 1).}$ In a similar
manner we obtain Ramond primaries
$$
\sum_{m'=0}^1X_2^{(2)}(m')V_{\ \ \ (1,2)}^{(1)0,m'}(z)S_{\ \ \
p,2,q}^{(k)mn}(z)
=W_{(p,s,q)}^{mn}(z),
\ \
s=\textstyle{1\over 2}(p+q\pm 1).\eqno(24)
$$
\par
$\underline{3.\ p+q\in 2{\bf Z},\ s_\pm={1\over 2}(p+q)\pm 1.}$
This case is a bit more complex, because
$$
\Delta_{(p,s_-,q)}=\Delta_{(p,s_+,q)},
$$
and action of $L_0^S=\oint_{C_z}{du\over 2 \pi i}uT_k^S(u)$
mixes $\phi_{(p,s_-,q)}$ and $\phi_{(p,s_+,q)}$. Finding
eigenvalues and eigenvectors of the `Hamiltonian matrix' we obtain
two kinds of fields: Neveu-Schwarz primaries
$$\eqalign{
\phi_{(2,1)}^{(1)\prime}(z)S_{\ \ \ pp'q}^{(k)mn}(z)&=
\sqrt{\textstyle{1\over 2}+y}
\ W_{(p,s_+,q)}^{mn}(z)-\sqrt{\textstyle{1\over 2}-y}
\ W_{(p,s_-,q)}^{mn}(z),
\cr
&y=[(k+4)p-(k+2)q]^{-1}
}\eqno(25)
$$
and Neveu-Schwarz superpartners
$$\eqalign{
S_{\ \ \ pp'q}^{(k)mn}(z)&=\sqrt{\textstyle{1\over 2}-y}
\ W_{(p,s_+,q)}^{mn}(z)+\sqrt{\textstyle{1\over 2}+y}
\ W_{(p,s_-,q)}^{mn}(z),
\cr
&s_\pm=\textstyle{1\over 2}(p+q)\pm 1,\ \ p'-3=p-q\ ({\rm mod}\ 4).
}\eqno(26)
$$
\par
Eqs. (23$-$26) allow to calculate structure
constants $C_{abc}^{S(k)}$, i.e. coefficients in tree-point
correlation functions of invariant primary fields $a$, $b$ and $c$.
Indeed, it is easy to check that, if we glue together two
vertex operators in the above-mentionned manner, the structure constants
of corresponding invariant fields are multiplied. For three
Neveu-Schwarz primaries or one Neveu-Schwarz and two Ramond primaries
we obtain
$$
C_{(p_1p'_1q_1)(p_2p'_2q_2)(p_3p'_3q_3)}^{S(k)}
=C_{(p_1,s_1)(p_2,s_2)(p_3,s_3)}^{(k)}
C_{(s_1,q_1)(s_2,q_2)(s_3,q_3)}^{(k+1)},\eqno(27)
$$
where $s_i={1\over 2}(p_i+q_i)$ if $p_i+q_i\in 2{\bf Z}$,
and $s_i={1\over 2}(p_i+q_i\pm 1)$
if $p_i+q_i\in 2{\bf Z}+1$; $C_{(p_1,s_1)(p_2,s_2)(p_3,s_3)}^{(k)}$
are well known structure constants of minimal model
M$_k$.$^9$ For calculations with superpartners it is necessary
to take into account the linear combination from Eq. (26).
\medskip
\par\noindent
{\bf Acknowlegements}
\medskip
\par\noindent
Author is grateful to S. Kryukov, Ya. P. Pugay, and S. Ye. Parkhomenko
for discussion.
\medskip
\par\noindent
{\bf References}
\par\noindent
1. A. B. Zamolodchikov, $Teor.$ $Mat.$ $Fiz.$ {\bf 65} (1985) 347
\par\noindent
2. H. Eichenherr, $Phys$. $Lett$. {\bf B151} (1985) 26
\par\noindent
3. M. A. Bershadsky, V. G. Knizhnik and M. G. Teitelman, $Phys$. $Lett$.
{\bf B151} (1985) 31
\par\noindent
4. D. Freedan, Z. Qiu and S. Shenker, $Phys$. $Lett$. {\bf B151} (1985) 37
\par\noindent
5. P. Goddard, A. Kent and D. Olive, $Commun$. $Math$. $Phys$.
{\bf 103} (1986) 105
\par\noindent
6. V. G. Knizhnik and A. B. Zamolodchikov, $Nucl$. $Phys$. {\bf B247} (1984) 83
\par\noindent
7. Vl. S. Dotsenko and V. A. Fateev, $Nucl$. $Phys$.
{\bf B240 [FS12]} (1984) 312
\par\noindent
8. Vl. S. Dotsenko and V. A. Fateev, $Nucl$. $Phys$.
{\bf B251 [FS13]} (1985) 691
\par\noindent
9. Vl. S. Dotsenko and V. A. Fateev, $Phys$. $Lett$. {\bf B154} (1985) 291
\par\noindent
10. G. Felder, $Nucl$. $Phys$. {\bf B317} (1989) 215
\par\noindent
11. C. Gomez and G. Sierra, $Nucl$. $Phys$. {\bf B352} (1991) 791
\par\noindent
12. J.-L. Gervais, preprint LPTENS-91/22, hep-th@xxx/9205034,
April 1992
\par\noindent
13. A. A. Belavin, A. M. Polyakov and A. B. Zamolodchikov, $Nucl$. $Phys$.
{\bf B241} (1984) 333
\par\noindent
14. M. Yu. Lashkevich, $Int$. $J$. $Mod$. $Phys$. {\bf A7} (1992) 6623
\par\noindent
15. M. Yu. Lashkevich, to be published in $Mod$. $Phys$. $Lett.$ $A$
\par\noindent
16. Vl. S. Dotsenko, $Adv$. $Stud$. $Pure$ $Math$. {\bf 16} (1988) 123
\par\noindent
17. A. B. Zamolodchikov and V. A. Fateev,
$Sov$. $Phys$. $JETP$ {\bf 62} (1985) 215
\par\noindent
18. M. Yu. Lashkevich, preprint LANDAU-92-TMP-1, October 1992
\end